\documentclass{article}
\usepackage{amssymb}

\input{tcilatex}
\begin{document}

\begin{center}
\bigskip \textbf{QUANTUM\ PHASE\ SHIFT IN CHERN-SIMONS MODIFIED GRAVITY}

\bigskip

\bigskip

\bigskip

K.K. Nandi$^{1,2,3,a}$, I.R. Kizirgulov$^{3,b}$, O.V. Mikolaychuk$^{3,c}$,
N.P. Mikolaychuk$^{3,d}$, A.A. Potapov$^{3,e}$

$^{1}$Department of Mathematics, University of North Bengal, Siliguri
734013, India

$^{2}$Joint Research Laboratory, Bashkir State Pedagogical University, Ufa
450000, Russia

$^{3}$Department of Theoretical Physics, Sterlitamak State Pedagogical
Academy, Sterlitamak 453103, Russia

$\bigskip $

$^{a}$Email: kamalnandi1952@yahoo.co.in

$^{b}$Email: kizirgulovir@mail.ru

$^{c}$Email: lacertida@mail.ru

$^{d}$Email: zetavoznichego@mail.ru

$^{e}$Email: potapovaa2008@rambler.ru

\bigskip

\textbf{Abstract}
\end{center}

Using a unified approach of optical-mechanical analogy in a semiclassical
formula, we evaluate the effect of Chern-Simons modified gravity on the
quantum phase shift of de Broglie waves in neutron interferometry. The phase
shift calculated here reveals, in a single equation, a combination of
effects coming from Newtonian gravity, inertial forces, Schwarzschild and
Chern-Simons modified gravity. However the last two effects, though \textit{%
new}, turn out to be too tiny to be observed, and hence only of academic
interest at present. The approximations, wherever used, as well as the
drawbacks of the non-dynamical approach are clearly indicated.

\bigskip

PACS numbers(s): 95.85.Ry, 12.15.Ff, 04.50.+h

\begin{center}
\bigskip

\textbf{I. Introduction}
\end{center}

Investigations at a semiclassical level of gravitational effects on quantum
interference are expected to be useful for the development of a theory of
quantum gravity [1]. Although the full theory is yet to appear, outcomes of
early seminal experiments by Colella, Overhauser and Werner (COW) [2] and an
improved later experiment by Werner, Staudenmann and Colella (WSC) [3],
measuring respectively the effects of Newtonian gravity and Coriolis force
on the phase shift of neutrons, continue to provide essential resources in
this direction. The effect of Coriolis force was first suggested by Page
[4]. Several other related works have also contributed useful information.
For instance, a semiclassical treatment reveals that quantum uncertainty in
the source variables induces uncertainties in the metric components of
gravity in a specific manner [5]. Another important phenomenon is the
neutrino flavor oscillation induced by gravity [6-9]. A different kind of
theoretical approach in the calculations of the phase of quantum particles,
neutrinos included, has yielded a very interesting result: The Dirac spin
1/2 particle has the exact covariant Stodolsky [10] phase $S/\hbar =(1/\hbar
)\int {p_{\mu }}dx^{\mu }$ in a static gravity field [11]. Some of these
investigations could provide appropriate theoretical backgrounds for
astrophysical applications, e.g., in the atmospheric neutrino experiments
[12], or in the observations involving $\gamma $-ray bursts [13].

On the other hand, the developments in matterwave interferometry have shown
a greater promise (over photon interferometry) in the measurements of the
phase shift due to Earth's gravity and axial rotation. In the absence of
external forces, the interference with itself of the de Broglie wave
associated with an ensemble of particles allows us to predict, via Huygen's
principle, the motion of the wave. The presence of an external field
modifies the motion causing a shift in the interference fringes. A classic
example is the Aharonov-Bohm (AB) fringe shift which provides information as
to how the motion of electrons are modified in the presence of a magnetic
potential. However, the situation becomes more complicated when one
introduces the effects of gravity and rotation on matterwave interferometry.

An intuitive and elegant special relativistic treatment of the quantum phase
shift for thermal neutrons has been proposed by Anandan [14]. However, the
effects of gravity and rotation are still considered as separate components.
On the other hand, a more comprehensive analysis involving gravity and
inertial forces would require the use of complete geometrical framework of
curved spacetime. The special relativistic effects should follow only as a
limiting case. Anandan's semiclassical treatment paves the way for the
application of simpler \textquotedblleft Newtonian" form of equations of
optical-mechanical analogy in curved spacetime [15-19]. These equations
provide a substitute for geometrical framework though only for calculational
purposes. Using them, the corrections to the phase shift had been worked out
[20] in certain spacetimes: A \textquotedblleft Coriolis" force leading to a
gravitational analog of the quantum AB effect appeared \textit{naturally}
from the application of a \textquotedblleft Lorentz force\textquotedblright\
form\ of the geodesic equation in Kerr and Kerr-Sen [21] spacetime. (Optical
analogs are artifacts designated by \textquotedblleft ..\textquotedblright\
and they can be easily translated into actual mechanical quantities, see
below.)

A different genre of theory that leads to rotating solution different from
that of Kerr is Chern-Simons (hereafter CS) modified gravity. A reasonable
summary of the timeline and study of CS modified gravity is in order. The CS
modified gravity descends from all versions of string theory upon
4-dimensional compactification due to the Green-Schwarz anomaly-cancelling
mechanism [21,22]. The first group to study CS modified gravity as a string
inspired correction to general relativity consisted of Alexander, Peshkin
and Sheikh-Jabbari [23] as well as Alexander and Gates [24]. The initial
non-dynamical formulation of CS\ modified gravity was presented by Jackiw
and Pi [25] who also showed that the Schwarzschild metric as well as a
special type of gravitational wave is a solution of CS modified gravity.
Guarrera and Hariton [26] showed, using Jackiw-Pi CS coupling field, that
Reissner-Nordstr\"{o}m is also a solution of CS gravity. The first spinning
black hole solutions were found by Alexander and Yunes [27, 28]. These
solutions represent spinning black holes in the far field limit but are not
axisymmetric. Smith, Erickcek, Caldwell, and Kamionkowski [29] evaluated
constraints on the CS parameter space from current satellite experiments
such as LAGEOS and Gravity Probe-B. Their approach can be regarded as an
extension of the Jackiw-Pi formulation because the action contains a scalar
field that is time varying but spatially homogeneous.\ Grumiller and Yunes
[30] performed a detailed study of exact solutions that could represent
spinning black holes with arbitrary CS coupling fields. This is the most
complete search for black hole-like solutions currently available in CS
modified gravity for a non-dynamical field. Alexander and Yunes [31] studied
the coupling of fermions to CS modified gravity. Yunes and Sopuerta [32]
studied perturbations of Schwarzschild black holes in CS modified gravity.

Konno, Matsuyama, Asano and Tanda [33] (hereafter KMAT) studied a solution
for a slowly rotating black hole in CS modified gravity derived previously
by Konno, Matsuyama and Tanda [34] (hereafter KMT). The KMT slow rotating
solution corresponds to a (special) spatial coupling field. The solution
differs from all others in that it is axisymmetric, among other things.
However, the solution and the non-dynamical approach have several unphysical
features. The solution is based on a purely \textit{ad hoc} non-dynamical
scalar, which plagues the well-posedness of the initial value problem [32].
Moreover, the solution is non-unique and does not carry over to the
dynamical sector because the scalar field has infinite energy, which in turn
prevents a self-consistent perturbation of the general relativity solutions
[35]. Despite these drawbacks, the Alexander-Yunes far field solution and
the KMT slow rotating solution are useful for understanding the CS effects
on physical observables. KMAT [33] showed that their solution remarkably
explains galactic flat rotation curves without the need to invoke
hypothetical dark matter. They also calculated the effect of frame dragging
on spin precession, which could provide a way to observationally test CS
gravity.

In this paper, we shall consider the KMT solution [33,34] to study the
effects of CS modified gravity on the phase shift in neutron interferometry
using, in addition, the equations of optical-mechanical analogy for rotating
sources [36]. The final formula for phase shift shows a combination of
different effects. We also recover, as a limiting case, the observed zeroth
order effects of Newtonian gravity and Coriolis force. As shown by KMAT, one
of the CS\ constants appear as constant tangential velocity in the galactic
halo. Thus one might treat its appearance in the phase shift as a non-local
effect similar to the effect of, say, cosmological constant in local
experiments. Our calculation shows that the effect of CS constants, which
represents dragging of inertial frames, appear only in the second order of
slow rotation parameter and thus too tiny.

The paper is organized as follows: To be reasonably informative, we briefly
state in Sec.II the CS action together with the KMT solution and discuss
stability of circular orbits, which is of crucial importance. The involved
approximations on the geodesic equation are clearly mentioned. Sec.III
describes the essence of Anandan's argument which provides a core \ element
of the ensuing analysis. To familiarize the readers at large with what we
refer to as optical mechanical analogy in curved spacetime, we briefly
review its contents in Sec.IV. We calculate the exact phase shift in CS
gravity in Sec.V and extract the corrections in Sec.VI. In Sec.VII, we
summarize the results of the paper together with a brief review of the
difficulties of the KMT solution and the non-dynamical approach. We shall
take $G=c_{0}=1$, unless specifically displayed.

\begin{center}
\textbf{II. The CS action, KMT solution and stability}
\end{center}

(a) CS modified gravity follows from the action [33,34] 
\begin{equation}
I=\!\frac{1}{16\pi }\int d^{4}x\left( \sqrt{-g}R-\frac{1}{2}v_{\mu }K^{\mu
}\right) ,
\end{equation}%
where $v_{\mu }\equiv \partial _{\mu }\vartheta $ is an external 4-vector,
the so-called the embedding coordinate. The CS topological current $K^{\mu }$
is given by 
\begin{equation}
K^{\mu }=\varepsilon ^{\mu \alpha \beta \gamma }\left[ \Gamma _{\alpha \tau
}^{\sigma }\partial _{\beta }\Gamma _{\gamma \sigma }^{\tau }+\frac{2}{3}%
\Gamma _{\alpha \tau }^{\sigma }\Gamma _{\beta \eta }^{\tau }\Gamma _{\gamma
\sigma }^{\eta }\right] .
\end{equation}%
where $\varepsilon ^{\sigma \mu \alpha \beta }$ the Levi-Civita tensor
density of weight one. Variation of the action with respect to the metric $%
g_{\mu \nu }$ leads to the CS field equation 
\begin{equation}
G^{\mu \nu }+C^{\mu \nu }=-8\pi T^{\mu \nu },
\end{equation}%
where $G^{\mu \nu }\equiv R^{\mu \nu }-\frac{1}{2}g^{\mu \nu }R$ is the
Einstein tensor, $T^{\mu \nu }$ is the usual energy-momentum tensor, and $%
C^{\mu \nu }$ is the Cotton tensor given by 
\begin{equation}
C^{\mu \nu }=-\frac{1}{2\sqrt{-g}}\left[ v_{\sigma }\left( \varepsilon
^{\sigma \mu \alpha \beta }\nabla _{\alpha }R_{\ \beta }^{\nu }+\varepsilon
^{\sigma \nu \alpha \beta }\nabla _{\alpha }R_{\ \beta }^{\mu }\right)
+\left( \nabla _{\sigma }v_{\tau }\right) \left( \!^{\ast }\!R^{\tau \mu
\sigma \nu }+\!^{\ast }\!R^{\tau \nu \sigma \mu }\right) \right]
\end{equation}%
In the above, the dual Riemann tensor is defined by $^{\ast }\!R_{\ \sigma
}^{\tau \ \mu \nu }\equiv \frac{1}{2}\varepsilon ^{\mu \nu \alpha \beta
}R_{\ \sigma \alpha \beta }^{\tau }$ and $v_{\tau \sigma }\equiv \nabla
_{\sigma }v_{\tau }=\partial _{\sigma }\partial _{\tau }\vartheta -\Gamma
_{\tau \sigma }^{\lambda }\partial _{\lambda }\vartheta $ is a symmetric
tensor. The Bianchi identity $\nabla _{\mu }G^{\mu \nu }=0$ and the equation
of motion $\nabla _{\mu }T^{\mu \nu }=0$ together imposes the condition
[33,34] 
\begin{equation}
0=\nabla _{\mu }C^{\mu \nu }=\frac{1}{8\sqrt{-g}}v^{\nu }\;\!^{\ast }\!R_{\
\tau }^{\sigma \ \mu \lambda }R_{\ \sigma \mu \lambda }^{\tau }.
\end{equation}

(b) KMAT [33] consider perturbation around the Schwarzschild solution of
mass $M$ in the small expansion parameter $\epsilon (\equiv J/Mr)$ where $J$
is the angular momentum. Then, for a slowly rotating star, the KMT solution
[34] in the case of a spacelike vector $v_{\mu }=\partial _{\mu }\vartheta
=\partial _{\mu }z=\partial _{\mu }\left( r\cos \theta \right) =\left(
0,\cos \theta ,-r\sin \theta ,0\right) $ becomes 
\begin{equation}
ds^{2}=-\left( 1-\frac{2M}{r}\right) dt^{2}+\left( 1-\frac{2M}{r}\right)
^{-1}dr^{2}+r^{2}\left( d\theta ^{2}+\sin ^{2}\theta d\phi ^{2}\right)
-2r^{2}\omega (r)dtd\phi ,
\end{equation}%
the $\theta $-dependence of the $(t\phi )$-component is restricted by
Eq.~(5), and 
\begin{equation}
\omega (r)=\frac{C_{1}}{r^{2}}\left( 1-\frac{2M}{r}\right) +\frac{C_{2}}{%
r^{3}}[r^{2}-2Mr-4M^{2}+4M(r-2M)\ln (r-2M)]
\end{equation}%
where $C_{1}$ and $C_{2}$ are constants (hereafter called CS constants)
characterizing slow rotation i.e., $C_{1},C_{2}\sim O(\epsilon )$. From the
metric, it is clear that $\omega \sim $ (length)$^{-1}$, which implies that $%
C_{1}\sim $ (length) and $C_{2}$ is truly dimensionless.

We should clearly mention that we are considering only the non-dynamical
formulation of CS\ modified gravity, where the CS scalar field in the KMT
solution satisfies the evolution equation only to leading order in the spin
[35] (See Ref. [29] for the dynamical formulation). However, it is well to
remember that the non-dynamical theory have several physical drawbacks as
mentioned in the introduction. The solution (6) is based on the choice of a
particular non-dynamical scalar defined by $\vartheta =r\cos \theta /\lambda
_{0}$, where $\lambda _{0}$ is a constant. On the other hand, in the
non-dynamical framework, there are no well-motivated physical reasons for
particular choices of the scalar field except for simplifying the equations.
Furthermore, for a given choice of CS scalar the Pontryagin constraint
significantly restricts the class of allowed solutions, even to the point
where the non-dynamical theory may be over-constrained and lack a well-posed
initial value problem [32]. The hope to imbed the KMT solution in the
dynamical sector is dashed because the stress energy associated with $%
\vartheta =r\cos \theta /\lambda _{0}$ is infinite. As shown by Yunes and
Pretorius [35], even a more general choice of the KMT scalar does not heal
the disease. Therefore the KMT metric is not a self-consistent solution to
the dynamical field equations. Yunes and Pretorius [35] have found a new
stationary axisymmetric solution for a different non-dynamical scalar.
Although the canonical $\vartheta $ is not allowed by this family of
solutions, it does bypass the problem with the stress tensor and is
compatible with the dynamical framework. The existence of new independent
solution shows that a solution in the non-dynamical theory such as that of
KMT can not be a unique black hole solution.

In the CS modified theory, as well as in general relativity, the standard
geodesic description holds if and only if the back-reaction is neglected in
the equations of motion. Such an approximation is valid, for example, in the
extreme-mass ratio limit. It is not valid, for instance, in the strong field
of equal mass binaries. Only under the assumption that the effect of test
particle's self-force on its own trajectory be neglected, the geodesic
calculations lead to the transverse circular velocity $v^{\phi }$ in the
equatorial plane ($\theta =\pi /2$) as 
\begin{equation}
v^{\phi }=r\frac{d\phi }{dt}=\pm \sqrt{\frac{M}{r}}+\left[ r\omega (r)+\frac{%
r^{2}}{2}\omega ^{\prime }(r)\right]
\end{equation}%
where $\omega ^{\prime }=d\omega /dr$. The first term clearly comes from the
Schwarzschild metric while the second term becomes a constant at large $r$,
i.e., 
\begin{equation}
v^{\phi }\simeq \pm \sqrt{\frac{M}{r}}+\frac{C_{2}}{2}.
\end{equation}%
This result remarkably explains the observed flat rotation curves in the
galactic halo. Essentially $C_{2}$ is responsible for frame dragging that
decays as $r^{-4}$.

(c)\ It is however not \textit{a priori }evident if the circular orbits in
the spacetime are stable. On the other hand, stability of circular orbits in
the galactic halo is a crucial requirement if any model of the halo has to
be physically viable. We have carried out the necessary calculations with
the potential 
\begin{equation}
V(r)=-\left( 1-\frac{2M}{r}\right) \left( 1+\frac{L^{2}}{r^{2}}\right)
-2EL\omega (r),
\end{equation}%
where the constants $E$ and $L$ are respectively the conserved relativistic
energy and angular momentum per unit rest mass of the test particle in a
circular orbit. Putting in the relevant expressions [33] for $E$ and $L$ in $%
\frac{d^{2}V}{dr^{2}}\mid _{r=R}$, we find the following results: The
expression for $V^{\prime \prime }=\frac{d^{2}V}{dr^{2}}\mid _{r=R}$ at any
arbitrary but fixed radius $r=R$ depends only on $C_{1}$ and $C_{2}$, and
that $\frac{d^{2}V}{dr^{2}}\mid _{r=R}<0$ only for values of these constants
very small relative to the length scale $M$, for instance $C_{1},C_{2}\in
\lbrack 0.001,0.0002]$ (see Fig.1 for details), which of course is in
perfect accordance with the slow rotation approximation. So we conclude that
the orbits are indeed stable for very small values of $C_{1}$, $C_{2}$.
Otherwise, the orbits are unstable.

\begin{center}
\textbf{III. Special relativistic phase shift: Anandan's formula}
\end{center}

Before we develop the equations for CS gravity, it is useful to have a brief
preview of the special relativistic phase shift because it will ultimately
provide the limiting case. (All special relativistic quantities will be
denoted by tilde). Using intuitive arguments, Anandan [14] proposed a
general formula for the quantum phase shift based on a correspondence
between the shift and the classical special relativistic equations of
motion. Consider two de Broglie wavelets originating at A and interfering at
B, one travelling along ADB and the other along ACB, the plane ADBC being
rotated by $180^{0}$ about the horizontal axis AO (Fig.2) while the Earth
itself is rotating about its axis with angular velocity $\vec{\Omega}$. The
vertical direction is determined by the resultant of gravity and the
centrifugal force of Earth. In the absence of any external forces, BD=BC and
we take CD = $d$, and AB = $l$. The external fields cause a shift in the
angular position of B by $\delta \theta \neq 0$ and thereby cause a phase
shift. Let $\tilde{v}$ be the velocity of the classical particle and ${%
\delta \tilde{v}}$ be the change in the velocity in the plane of
interference. Since $\delta \theta $ is small, $\sin \delta \theta \simeq
\delta \theta \simeq \frac{{\delta \tilde{v}}}{{\tilde{v}}}$. Then the
special relativistic phase shift $\Delta \phi _{SR}$ can be calculated as
[14]:%
\begin{equation}
\Delta \phi _{SR}=\tilde{\kappa}d\frac{{\delta \tilde{v}}}{{\tilde{v}}}=%
\frac{{\tilde{\kappa}d}}{{\tilde{v}}}\frac{{d\tilde{v}_{\bot }}}{{dt}}\frac{l%
}{{\tilde{v}}}=\frac{{\tilde{p}A}}{{\hbar \tilde{v}^{2}}}\frac{{d\tilde{v}%
_{\bot }}}{{dt}}=\frac{{\tilde{\gamma}mAd\tilde{v}_{\bot }/dt}}{{\tilde{v}%
\hbar }},
\end{equation}%
where $\tilde{\gamma}=\left( {1-\tilde{v}^{2}/c_{0}^{2}}\right) ^{-1/2}$, $%
A=ld$ is the planar area (assumed small) enclosed by the two paths of
interfering beams, $m$ is the neutron mass, $\tilde{p}(=\hbar \tilde{\kappa}%
) $ is the momentum in flat space, and $d\tilde{v}_{\bot }/dt$ is the
component of acceleration in the vertical direction CD. This remarkable
general formula, Eq.(11), at once gives rise to several individual
components.

The following cases were considered: (i) A particle with charge $e$ moving
in a magnetic field $B$. Then, $\tilde{\gamma}d\tilde{v}_{\bot }/dt=eB\tilde{%
v}/m$, which, when used in Eq.(11), immediately yields the Aharonov-Bohm
(AB) effect, viz., 
\begin{equation}
\Delta \widetilde{\phi }_{AB}=eBA/\hbar .
\end{equation}%
(ii) A spinless particle of mass $m$ in a gravitational field, $d\tilde{v}%
_{\bot }/dt=-g$, where $g$ is the gravitational acceleration on the surface
of the Earth. Using the expression for the relativistic momentum $\tilde{%
\gamma}^{2}=1+\tilde{p}^{2}/m^{2}c_{0}^{2}$ and the de Broglie relation $%
\tilde{p}=\hbar \tilde{\kappa}$, Eq.(11) leads to the following result: 
\begin{equation}
\Delta \widetilde{\phi }_{g}=-gAm^{2}/\hbar ^{2}\tilde{\kappa}-gA\tilde{%
\kappa}/c_{0}^{2}.
\end{equation}%
The COW experiment was accurate enough to measure the first quantum term,
but not the second, that is, the so called relativistic term. (iii) A
particle moving in the Coriolis force field of Earth so that, to first order
in $\Omega $, $d\tilde{v}_{\bot }/dt=2\left\vert \vec{\Omega}\times \vec{%
\tilde{v}}\right\vert =2\Omega _{n}\tilde{v}$, where $\Omega _{n}$ is the
component of Earth's angular velocity $\vec{\Omega}$ normal to the
apparatus. Using the Planck-Einstein law $E/c_{0}=\hbar \tilde{\omega}=mc_{0}%
\tilde{\gamma}$, one finds from Eq.(11) that 
\begin{equation}
\Delta \widetilde{\phi }_{cor}=\frac{2\tilde{\omega}A\Omega _{n}}{c_{0}}.
\end{equation}%
Using the dispersion relation%
\begin{equation}
\tilde{\omega}^{2}-\tilde{\kappa}^{2}=\frac{m^{2}c_{0}^{2}}{\hslash ^{2}}
\end{equation}%
we find from Eq.(14) that%
\[
\Delta \widetilde{\phi }_{cor}\simeq \frac{2mA\Omega _{n}}{\hslash }+\frac{%
\hslash A\Omega _{n}\tilde{\kappa}^{2}}{mc_{0}^{2}} 
\]%
The WSC experiment has tested the first nonrelativistic term to within a
good accuracy. (iv) This effect comes from the coupling of particle's spin
to the background curvature. Again using Eq.(11) together with the
Mathisson-Papapetrou force [37], the shift comes out to be 
\begin{equation}
\Delta \phi _{s.c.}=-\frac{{\hbar GMA\tilde{\omega}}}{{mc_{0}^{3}R^{3}}},
\end{equation}%
where $G$ is the Newtonian gravitational constant, $M$ is the gravitating
mass and $R$ is the distance from the center. (We shall not discuss this
spin-curvature component in this article.) With all the above in view, we
proceed to familiarize the readers with the salient features of our approach
in the curved spacetime regime.

\begin{center}
\textbf{IV.\ Optical-Mechanical equations in curved spacetime}
\end{center}

The central idea is to introduce an optical-mechanical analogy in curved
spacetime. The exact optical form of geodesic equations in the spherically
symmetric case of general relativity was originally derived in Ref.[15]. The
equations provide an excellent tool that enables one to visualize the
problems of geometrical optics as problems of classical mechanics and vice
versa. The first step is to find out an effective optical refractive index $%
n $ that is \textit{formally} equivalent to the geometrized gravity field.
This step in itself is not new. Usually the index equivalent to the exterior
Schwarzschild field is approximately taken to be $n\approx 1+2MG/rc_{0}^{2}$%
, where $M$ is the central gravitating mass. We derive the exact expression
as below.

Consider a static, spherically symmetric, but not necessarily vacuum,
solution of general relativity written in isotropic coordinates 
\begin{equation}
ds^{2}=h(\vec{r})c_{0}^{2}dt^{2}-\Phi ^{-2}(\vec{r})\left\vert {d\vec{r}}%
\right\vert ^{2},
\end{equation}%
where $\vec{r}\equiv (x,y,z)$ or $(r,\theta ,\varphi )$, and $h,\Phi $ could
be the solution of Einstein's field equations. Many metrics of physical
interest can be put into this isotropic form. The coordinate speed of light $%
c(\vec{r})$ is determined by the condition that the geodesic be null $\left(
ds^{2}=0\right) $: 
\begin{equation}
c(\vec{r})=\left\vert {\frac{{d\vec{r}}}{{dt}}}\right\vert =c_{0}\Phi \sqrt{h%
},
\end{equation}%
which immediately provides an effective index of refraction for light in the
gravitational field given by 
\begin{equation}
n(\vec{r})=\frac{1}{{\Phi \sqrt{h}}}.
\end{equation}%
The concept of optical mechanical analogy can be used to recast the geodesic
equation for \textit{both} massive and massless particles into a single,
exact Newtonian \textquotedblleft $\mathbf{F}=m\mathbf{a}$" type of equation
given by [16] 
\begin{equation}
\frac{{d^{2}\vec{r}}}{{dA^{2}}}=\vec{\nabla}\left( {\frac{1}{2}N^{2}c_{0}^{2}%
}\right) ,N(\vec{r})=n(\vec{r})\sqrt{1-\frac{{m^{2}c_{0}^{4}h}}{{E_{0}^{2}}}}%
,dA=\frac{{dt}}{{n^{2}}},
\end{equation}%
where $m$ is the rest mass of the particle, $E_{0}$ is the conserved total
energy, $\vec{\nabla}$ is the gradient operator. All the standard geodesic
equations in Schwarzschild gravity including some new insights in cosmology
follow from the above equation. This remarkably simple feature of the
geodesic equations is brought about by the use of the stepping parameter $A$%
, having dimension of length, first introduced by Evans and Rosenquist [19].
Eqs. (20) provide an easy way to introduce quantum relations so that the
geodesic motion of a particle can be interpreted as motion of de Broglie
matter waves in a dispersive medium with an effective index of refraction $N(%
\vec{r},\lambda )=n(\vec{r})/\sqrt{1+(\lambda /\lambda _{c})^{2}}$ where $%
\lambda _{c}$ is the Compton wavelength given by $\lambda _{c}=2\pi \hbar
/mc_{0}$ and $\lambda =\tilde{\lambda}\Phi ^{-1}$ is the physical wavelength
measured in a gravity field. This interpretation allows us to extend in a
straightforward manner the classical optical-mechanical analogy into the
quantum regime.

Alsing [36] has subsequently extended the method to broader class of metrics
in general relativity and his equation is going to provide the basic tool in
what follows. Consider the most general form of the metric given by 
\begin{equation}
ds^{2}=g_{00}c_{0}^{2}dt^{2}+2g_{0i}c_{0}dtdx^{i}+g_{ij}dx^{i}dx^{j},\quad
\quad i,j=1,2,3.
\end{equation}%
Define the proper time $d\tau $, proper length $dl$ and the velocity $v$
measured with respect to this proper time as 
\begin{equation}
v^{i}=\frac{{dx^{i}}}{{d\tau }},
\end{equation}%
\begin{equation}
\left. {\frac{{ds}}{{c_{0}}}}\right\vert _{v=0}=d\tau =\frac{\sqrt{h}}{{c_{0}%
}}(c_{0}dt-g_{i}dx^{i}),\quad \quad g_{00}=h,
\end{equation}%
\begin{equation}
dl^{2}=\gamma _{ij}dx^{i}dx^{j}=\left( {-g_{ij}+\frac{{g_{0i}g_{0j}}}{{g_{00}%
}}}\right) dx^{i}dx^{j},\quad \quad v^{2}=\gamma _{ij}v^{i}v^{j},
\end{equation}%
\begin{equation}
g_{i}=-\frac{{g_{0i}}}{{g_{00}}},\quad \quad g^{i}=\gamma ^{ij}g_{j}=-g^{0i}.
\end{equation}%
The angular momentum $g_{i}$ and the proper velocity $v_{i}$ are vectors
defined in the 3-space characterized by the metric $\gamma _{ij}$, which is
used to raise or lower the indices of these 3-vectors. Now, metric (21) can
be rewritten as 
\begin{equation}
ds^{2}=h(c_{0}dt-g_{i}dx^{i})^{2}\left( {1-v^{2}/c_{0}^{2}}\right) .
\end{equation}%
and the conserved energy $E$ is given by [38] 
\begin{equation}
E=mc_{0}^{2}g_{0\alpha }\frac{{dx^{\alpha }}}{{ds}}=\frac{{mc_{0}^{2}\sqrt{h}%
}}{\sqrt{1-v^{2}/c_{0}^{2}}},\quad \quad \alpha =0,1,2,3.
\end{equation}%
The variational principle for the geodesics following from Eq.(21) is given
by 
\begin{equation}
\delta \int\limits_{\vec{x}_{1},t_{1}}^{\vec{x}_{2},t_{2}}{mc_{0}ds=}\delta
\int\limits_{\vec{x}_{1},t_{1}}^{\vec{x}_{2},t_{2}}{Ldt=}\delta \int\limits_{%
\vec{x}_{1},t_{1}}^{\vec{x}_{2},t_{2}}{mc_{0}^{2}\sqrt{h(\vec{r})}\sqrt{%
1-v^{2}(\vec{r},\tilde{v})/c_{0}^{2}}}\tilde{\beta}(\vec{r},\tilde{v})dt=0,
\end{equation}%
where 
\begin{equation}
\tilde{\beta}=1-\frac{{g_{i}\tilde{v}^{i}}}{{c_{0}}},\quad \quad \tilde{v}%
^{i}=\frac{{dx^{i}}}{{dt}},\quad \quad v^{i}=\frac{{\tilde{v}^{i}}}{{\tilde{%
\beta}\sqrt{h}}}.
\end{equation}%
From the Lagrangian $L$, let us find the momenta conjugate to $\tilde{v}^{i}$%
. (Flat space quantities are designated by tilda). This is given by 
\begin{equation}
\frac{{\partial L}}{{\partial \tilde{v}^{i}}}=E\left( {g_{i}+\frac{{v_{i}}}{%
\sqrt{h}}}\right) .
\end{equation}%
Since the vectors $g_{i}$ and $v_{i}$ are defined in a space with the metric 
$\gamma _{ij}$, we can identify the right hand side of the Eq.(30) as a
vector in the same 3-space. Let us call it the momentum 3-vector 
\begin{equation}
p_{i}\equiv E\left( {g_{i}+\frac{{v_{i}}}{\sqrt{h}}}\right) .
\end{equation}%
It can also be verified that $H=\frac{{\partial L}}{{\partial \tilde{v}^{i}}}%
\tilde{v}^{i}-L\equiv p_{i}\tilde{v}^{i}-L=E$, which is a constant along the
trajectory of a particle as stated in Eq.(27). Hence, it is possible to
introduce Maupertuis principle $\delta \int\limits_{\vec{x}_{1}}^{\vec{x}%
_{2}}{p_{i}\tilde{v}^{i}}dt=0$. Assuming further that the spatial part of
the metric could be written in an isotropic form $dl=dl_{E}/\Phi
,dl_{E}=\delta _{ij}dx^{i}dx^{j},\gamma _{ij}=\delta _{ij}\Phi ^{-2}$, we get%
\begin{equation}
v^{2}=n^{2}\tilde{v}^{2}/\tilde{\beta}^{2},n=\frac{1}{{\Phi \sqrt{h}}}
\end{equation}%
and the Maupertuis variational principle yields, after lengthy algebra
introducing the parameter $A$, the geodesic equation in the form 
\begin{equation}
\frac{{d^{2}\vec{r}}}{{dA^{2}}}=\vec{\nabla}\left( {\frac{{n^{2}v^{2}}}{2}}%
\right) +\frac{{d\vec{r}}}{{dA}}\times (\vec{\nabla}\times \vec{g}),
\end{equation}%
\begin{equation}
dA=n^{-2}\tilde{\beta}dt,\frac{d\tau }{dA}=n^{2}\sqrt{h}
\end{equation}%
where $\vec{g}\equiv (g_{i})$. This Newtonian form of the geodesic equation,
valid for both massless and massive particles, has been obtained under the
only assumption that the spatial part of the metric could be written in an
isotropic form. On eliminating $A$ from Eqs.(33) and (34), we can find the
rotational contributions to the well known Schwarzschild orbits for light
and planets [36]. The geodesic Eqs.(33),(34) are valid only under the
assumption of small test body approximation (back reaction ignored, as
stated in Sec.\textbf{II}, before Eq.(8))

Eq.(33) admits an immediate interpretation, \textit{albeit }in optical
parlance, as describing the motion of a particle in a \textquotedblleft
potential" $(-1/2)n^{2}v^{2}$ and subjected to a \textquotedblleft Coriolis"
force $\frac{{d\vec{r}}}{{dA}}\times (\vec{\nabla}\times \vec{g})$, which
would appear, for instance, in the absence of gravity $(n=1,h=1)$ in a
coordinate system rotating with angular velocity $\overrightarrow{\Omega }%
=(1/2)(\vec{\nabla}\times \vec{g})$. One might look at the first term in
Eq.(33) as the \textquotedblleft gravistatic"\textit{\ }part and the second
term as the \textquotedblleft gravimagnetic" part. However, such
gravimagnetic analogy practically works out only in the weak field limit of
physically interesting metrics. The reason is that the coordinate
transformation needed to achieve the spatially isotropic form in general
entails use of implicit functions. To avoid this complication, one focuses
attention only to large distances on the equatorial plane (see Sec.\textbf{V}
below). It would be interesting to compare and contrast the full optical
analogy equations with the general relativistic mechanical equations
developed in Ref.[39]. The gravimagnetic Eqs.(33), (34) do yield the known
equations in the weak field, slow velocity limit [40]. In CS modified
gravity, the gravimagnetic analogy was first applied by Alexander and Yunes
[28] followed by Smith \textit{et al} [29].

If one considers that gravity is given by the nonrelativistic, Newtonian law 
$\ \tilde{v}^{2}=\frac{2M}{r}$ in an otherwise flat space $(n=1,h=1)$ so
that $dA=d\tau $, then one has from Eq.(33), the component of radial
acceleration 
\begin{equation}
\frac{{d\tilde{v}_{\bot }}}{{d\tau }}=-g+2\left\vert \vec{\tilde{v}}\times 
\vec{\Omega}\right\vert ,
\end{equation}%
which leads to 
\begin{equation}
\Delta \phi _{SR}=\left[ -g+2\left\vert \vec{\tilde{v}}\times \vec{\Omega}%
\right\vert \right] \left( \frac{\tilde{p}A}{\hbar \tilde{v}^{2}}\right) ,
\end{equation}%
where 
\begin{equation}
-g=\frac{1}{2}\frac{d}{{dr}}(\tilde{v}^{2})=-\frac{M}{r^{2}}.
\end{equation}%
One can see how beautifully the effects of gravity and rotation, otherwise
considered as separate components in the literature [cases (ii) and (iii) in
Sec.III], are synthesized into a single equation. We are now ready to apply
Alsing Eqs.(33), (34) for calculating the phase shift in CS gravity.

\begin{center}
\textbf{V. Phase shift in CS gravity}
\end{center}

To apply Eqs.(33), (34) we need to cast the KMT metric (6) in a spatially
isotropic form. In general this is not possible, so for practical reasons,
we restrict ourselves to equatorial plane ($\theta =\pi /2$) and to the weak
field regime of the rotating source, $r\gg 2M$. Neglecting $\omega ^{2}$, we
have 
\begin{equation}
ds^{2}=-h(r)\left[ dt+\frac{\omega r^{2}}{h(r)}d\phi \right] ^{2}+\left( 1-%
\frac{2M}{r}\right) ^{-1}dr^{2}+r^{2}d\phi ^{2},
\end{equation}%
\begin{equation}
h(r)=1-\frac{2M}{r},
\end{equation}%
\begin{equation}
g_{\phi }=-\frac{\omega (r)r^{2}}{h(r)}.
\end{equation}%
To achieve the isotropic form, we adopt isotropic radial variable $\rho $
via $\Phi =\frac{\rho }{r}$ and $\frac{d\rho }{\rho }=(1+M/r)(dr/r)$, and
obtain $\rho =re^{-\frac{M}{r}}$, which yields $r=\rho +M$, to lowest order
in $\frac{M}{r}$. Thus, in all, we have the expressions with $n$ given by
Eq.(32): 
\begin{equation}
\Phi (\rho )\simeq 1-\frac{M}{\rho },h(\rho )\simeq 1-\frac{2M}{\rho },
\end{equation}%
\begin{equation}
n(\rho )\simeq 1+\frac{2M}{\rho },g_{\phi }\simeq -\left( 1+\frac{2M}{\rho }%
\right) (\rho +M)^{2}\omega (\rho +M),
\end{equation}%
and the metric (38) for $\frac{M}{\rho }\ll 1$ takes the form 
\begin{equation}
ds^{2}=-\left( 1-\frac{2M}{\rho }\right) \left[ dt+g_{\phi }d\phi \right]
^{2}+\left( 1+\frac{2M}{\rho }\right) [d\rho ^{2}+\rho ^{2}d\phi ^{2}].
\end{equation}%
With these at hand, and using Eq.(33), we can rewrite the radial and cross
radial components of \textquotedblleft acceleration\textquotedblright\ as
follows: 
\begin{equation}
\frac{d^{2}\rho }{dA^{2}}-\rho \left( \frac{d\phi }{dA}\right) ^{2}=\frac{d}{%
d\rho }\left( \frac{n^{2}v^{2}}{2}\right) +\left\vert 2\frac{d%
\overrightarrow{\rho }}{dA}\times \overrightarrow{\Omega }\right\vert ,
\end{equation}%
\begin{equation}
\rho \frac{d^{2}\phi }{dA^{2}}+2\frac{d\rho }{dA}\frac{d\phi }{dA}=-\frac{1}{%
\rho }\frac{d\rho }{dA}\frac{dg_{\phi }}{d\rho },
\end{equation}%
where $\overrightarrow{\Omega }\equiv \frac{1}{2}(\overrightarrow{\nabla }%
\times \overrightarrow{g})$ is the angular velocity of rotation of the
gravitating mass $M$. The last term in Eq.(44) represents \textquotedblleft
Coriolis force". The first integral of Eq.(45) is%
\begin{equation}
\frac{d}{dA}\left( \rho ^{2}\frac{d\phi }{dA}+g_{\phi }\right) =0\Rightarrow
\rho ^{2}\frac{d\phi }{dA}+g_{\phi }\equiv \ell
\end{equation}%
where $\ell $ is an arbitrary constant of integration, which may be
interpreted as the conservation of total \textquotedblleft angular
momentum". Thus%
\begin{equation}
\frac{d\phi }{dA}=\frac{\ell -g_{\phi }}{\rho ^{2}}.
\end{equation}

\ So far we have been talking in the language of optical analogue
(\textquotedblleft ..") of mechanical quantities. Let us now translate them
into mechanical quantities proper, quantities that are actually measured in
a gravity field. Thus we calculate the radial acceleration $\frac{dv_{\perp }%
}{d\tau }$ as follows. Converting $A\rightarrow \tau $ by means of Eq.(34),
we can rewrite Eq.(44) as 
\begin{equation}
\frac{dv_{\perp }}{d\tau }\equiv \frac{d^{2}\rho }{d\tau ^{2}}-\rho \left( 
\frac{d\phi }{d\tau }\right) ^{2}
\end{equation}%
\begin{equation}
=\left( \frac{1}{n^{4}h}\right) \left[ \frac{d}{d\rho }\left( \frac{%
n^{2}v^{2}}{2}\right) +\left( n^{2}\sqrt{h}\right) (2\Omega _{n}v)\right]
\end{equation}%
where we have used in Eq.(44) the following: $\left\vert 2\frac{d%
\overrightarrow{\rho }}{dA}\times \overrightarrow{\Omega }\right\vert
=2\left( \frac{d\rho }{d\tau }\frac{d\tau }{dA}\right) \Omega _{n}$ together
with$\frac{d\rho }{d\tau }=v$, $\frac{d\tau }{dA}=n^{2}\sqrt{h}$, $\Omega
_{n}$ being the angular velocity of rotation of $M$ normal to the plane of
the interferometer. We now need to calculate $\widetilde{\beta }$. Using
Eqs.(29), (34) and Eq.(47) we find 
\begin{equation}
\tilde{\beta}=1-{g_{\phi }}\frac{d\phi }{dt}=1-{g_{\phi }}\frac{d\phi }{dA}%
\frac{dA}{dt}=1-g_{\phi }\left[ \frac{\ell -g_{\phi }}{n^{2}\rho ^{2}}\right]
\tilde{\beta}
\end{equation}%
Since we disregard terms of the order of $g_{\phi }^{2}$, being of the order
of $\omega ^{2}$, we end up with%
\begin{equation}
\tilde{\beta}^{-1}\simeq \left[ 1+\frac{g_{\phi }\ell }{n^{2}\rho ^{2}}%
\right] .
\end{equation}%
Then noting that $\tilde{v}^{2}=2M/r\simeq 2M/\rho $, and using $n$ from
Eqs.(42) and $\tilde{\beta}$ from Eq.(51), we can finally express $v$ of
Eq.(32) as%
\begin{equation}
v=\frac{n\widetilde{v}}{\widetilde{\beta }}=\left( 1+\frac{2M}{\rho }\right) 
\sqrt{\frac{2M}{\rho }}\left[ 1+\frac{g_{\phi }\ell }{n^{2}\rho ^{2}}\right]
.
\end{equation}%
On the other hand, we have the conserved proper energy and momentum for a
test particle of rest mass $m$ as follows [cf. Eq.(27)] 
\begin{equation}
E=\frac{m\sqrt{h}}{\sqrt{1-v^{2}}}=\widetilde{E}\sqrt{h}
\end{equation}%
\begin{equation}
p=\frac{mv}{\sqrt{1-v^{2}}}\Phi ^{-1}=\widetilde{p}\Phi ^{-1}
\end{equation}%
such that the mass-shell condition may be written in a special relativistic
form%
\begin{equation}
m^{2}=h\widetilde{E}^{2}-\Phi ^{-2}\widetilde{p}^{2}=E^{2}-p^{2}.
\end{equation}%
(Note that, we\ could obtain, alternatively, the refractive index $n$ by
putting $m=0$ so that $c=\frac{\widetilde{p}}{\widetilde{E}}=c_{0}\Phi \sqrt{%
h}=c_{0}n^{-1}$.) Using Eq.(54), we can express $v^{2}$ in terms of $p$ and $%
\Phi $, and further introducing the quantum relation $\widetilde{p}=\hslash 
\widetilde{\kappa }$, we get the curved space corrected Newtonian gravity
term, $\Delta \widetilde{\phi }_{g}\rightarrow \Delta \phi _{g}$: 
\begin{eqnarray}
\Delta \phi _{g} &=&-g\frac{pA}{\hslash v^{2}}\equiv -Ag\left( \frac{%
m^{2}c_{0}^{2}+p^{2}\Phi ^{2}}{\hslash c_{0}^{2}p\Phi ^{2}}\right)  \nonumber
\\
&=&-g\left( \frac{Am^{2}}{\hslash ^{2}\widetilde{\kappa }}\right) \Phi ^{-2}-%
\frac{gA\widetilde{\kappa }}{c_{0}^{2}} \\
&\simeq &-g\left( \frac{Am^{2}}{\hslash ^{2}\widetilde{\kappa }}\right)
\left( 1+\frac{2M}{\rho }\right) -\frac{gA\widetilde{\kappa }}{c_{0}^{2}}
\end{eqnarray}%
An interesting observation is that the relativistic second term is
completely unaffected by gravity. Symbolically, we write the phase shift in
CS gravity as 
\begin{eqnarray}
\Delta \phi _{CSG} &=&\left( \frac{dv_{\perp }}{d\tau }\right) \times \left( 
\frac{pA}{\hslash v^{2}}\right) \\
&=&\left( \frac{1}{n^{4}h}\right) \left[ \frac{d}{d\rho }\left( \frac{%
n^{2}v^{2}}{2}\right) +\left( n^{2}\sqrt{h}\right) (2\Omega _{n}v)\right]
\times \left( \frac{pA}{\hslash v^{2}}\right) \\
&=&\Delta \phi _{1}+\Delta \phi _{2},
\end{eqnarray}%
where $\Delta \phi _{1}$ and $\Delta \phi _{2}$ are gravitational and
Coriolis contributions respectively. This is the exact expression that
contains terms of all orders in ($\frac{1}{\rho }$) and higher. All we have
to do now is to collect the expressions for $\Phi $, $h$, $n$, $g_{\varphi }$%
, $v$ and $\frac{dv_{\perp }}{d\tau }$ from Eqs. (41), (42), (49) and (52),
insert them in Eq.(60), expand and find contributions to lowest order in ($%
\frac{1}{\rho }$).

\begin{center}
\textbf{VI. Lowest order terms}
\end{center}

To extract ostensibly first order terms, we ignore second and higher order
terms in $M$, $C_{1}$, $C_{2}$ together with their various products. Then
the shift $\Delta \phi _{1}$ to lowest order is as follows: 
\begin{eqnarray}
\Delta \phi _{1} &=&\left( \frac{1}{n^{4}h}\right) \left( \frac{pA}{\hslash
v^{2}}\right) \frac{d}{d\rho }\left( \frac{n^{2}v^{2}}{2}\right) \\
&\simeq &\left[ 1+\ell \left\{ \frac{4C_{2}}{\rho }+\frac{6C_{1}}{\rho ^{2}}%
\right\} +O(\epsilon ^{2})\right] \left( \Delta \phi _{g}\right)
\end{eqnarray}%
We can see how the term $\Delta \phi _{1}$, which represents only $\Delta 
\widetilde{\phi }_{g}$ in flat space, is modified by the quantities $C_{1}$, 
$C_{2}$ and $\frac{M}{\rho }$.

To calculate the Coriolis effect, we introduce the quantum relation $%
\widetilde{E}=\hslash \widetilde{\omega }$ so that 
\begin{eqnarray}
\left( \frac{pA}{\hslash v^{2}}\right) &=&\frac{AE\Phi ^{-1}}{\hslash v\sqrt{%
h}}=\frac{A\widetilde{\omega }\Phi ^{-1}}{v} \\
&\Rightarrow &(2\Omega _{n}v)\left( \frac{pA}{\hslash v^{2}}\right) =2%
\widetilde{\omega }\Omega _{n}A\Phi ^{-1}.
\end{eqnarray}%
Thus we get from Eqs.(60) and (64) 
\begin{equation}
\Delta \phi _{2}=\left[ \left( n^{2}\sqrt{h}\right) ^{-1}\right] \Phi ^{-1}(2%
\widetilde{\omega }\Omega _{n}A)\equiv \left( 1+\frac{4M}{\rho }\right) (2%
\widetilde{\omega }\Omega _{n}A).
\end{equation}%
Using the dispersion relation from Eq.(55)%
\begin{equation}
\widetilde{\omega }^{2}=\frac{m^{2}}{\hslash ^{2}}\left[ \frac{1}{h(\rho )}+%
\frac{\Phi ^{-2}}{h(\rho )}\left( \frac{\hslash ^{2}\widetilde{\kappa }^{2}}{%
m^{2}}\right) \right]
\end{equation}%
in $2\widetilde{\omega }\Omega _{n}A$, we explicitly obtain the gravity
corrected Coriolis term, $\Delta \widetilde{\phi }_{cor}\rightarrow \Delta
\phi _{cor}$: 
\begin{eqnarray}
\Delta \phi _{cor} &=&2\widetilde{\omega }\Omega _{n}A \\
&\simeq &\left( 1+\frac{M}{\rho }\right) \left[ \frac{2A\Omega _{n}m}{%
\hslash }\right] +\left( 1+\frac{3M}{\rho }\right) \left[ \frac{\hslash 
\widetilde{\kappa }^{2}A\Omega _{n}}{mc_{0}^{2}}\right] .
\end{eqnarray}%
Thus we have, from Eq.(65), 
\[
\Delta \phi _{2}=\left( 1+\frac{4M}{\rho }\right) (\Delta \phi _{cor}). 
\]%
Recombining it with Eq.(62), we get 
\begin{equation}
\Delta \phi _{CSG}\simeq \left[ 1+\ell \left\{ \frac{4C_{2}}{\rho }+\frac{%
6C_{1}}{\rho ^{2}}\right\} \right] \left( \Delta \phi _{g}\right) +\left( 1+%
\frac{4M}{\rho }\right) (\Delta \phi _{cor}).
\end{equation}%
Recall that $C_{1}$ has the dimension of length, $C_{2}$ is dimensionless
and $\ell $ has the dimension of length, hence the CS effect $\ell \left\{ 
\frac{4C_{2}}{\rho }+\frac{6C_{1}}{\rho ^{2}}\right\} $ is dimensionless, as
should be the case. Eq.(69) is the main result of our paper.

As we see, the effect is coupled with $\ell $ which is a sum of
\textquotedblleft angular momentum\textquotedblright\ $\rho ^{2}\frac{d\phi 
}{dA}$ of particles and $g_{\phi }$ of the source. The interferometer plane
is usually placed at two vertical orientations so that the angular momentum
of particles moving in the plane is practically zero. We are then left with
only the source influence $g_{\phi }$, which, when multiplied by $C$,
becomes an effect in $O(\epsilon ^{2})$. Nevertheless, a number of
observations can be made. It is immediate that in the limit of flat space ($%
M=0$, $C_{1}$, $C_{2}=0$), we recover the individual special relativistic
components $\Delta \widetilde{\phi }_{g}$ and $\Delta \widetilde{\phi }%
_{cor} $ derived by Anandan [14]. This means, in this limit, $\Delta \phi
_{CSG}=\Delta \phi _{SR}$. If $C_{1}$, $C_{2}=0$, but $M\neq 0$, we recover
the influence of Schwarzschild gravity embodied in the factors of $\left( 
\frac{M}{\rho }\right) $, so that $\Delta \phi _{CSG}=\Delta \phi _{Sch}$.
It is a new result by itself. If the black hole mass $M=0$ but $C_{1}$, $%
C_{2}$ $\neq 0$, we are left with the effect of a mere rotating coordinate
system, which is of no physical interest. The remarkable result is that the
parameters $C_{1}$, $C_{2}$ do\textit{\ not} affect the Coriolis term at
all, while it modifies only the pure gravity term $\Delta \phi _{g}$. This
is because $v$, the carrier of $C_{1}$, $C_{2}$ via $g_{\phi }$, cancels out
as shown in Eq.(64).

Since the CS component of the phase shift depends crucially on the factor $%
\ell $, let us analyze what different values of it means. Although it is an
optical constant, we see from Eq.(46) that it is also constant in proper
time, $\frac{d\ell }{d\tau }=0$, because $\frac{d\tau }{dA}\neq 0$. Note
further that $\ell $\textsc{\ }consists of two parts: A part $\rho ^{2}\frac{%
d\phi }{dA}$, proportional to the usual particle orbital angular momentum,
and a part $g_{\phi }$ proportional to the angular momentum of the black
hole itself. If $\ell =0$, the CS effect vanishes altogether. This is the
case when a radially freely falling particle acquires an angular velocity $%
\frac{d\phi }{dt}\varpropto \frac{g_{\phi }}{\rho ^{2}}$ near the source in
the \textit{same} sense as the rotation of the black hole, exactly similar
to what happens in the Lense-Thirring drag [41]. On the other hand, if $\ell
=g_{\phi }$, then $\rho ^{2}\frac{d\phi }{dA}=0\Rightarrow \frac{d\phi }{dt}%
=0$, that is, the particle is at rest at constant ($r,\theta $) relative to
distant stars, and to remain at rest it has to swim against the spacetime
drag. In this case, we see from Eq.(51) that there can be no $O(\epsilon )$
effect because the lowest order CS component would go like $g_{\phi }\times
C\sim O(\epsilon ^{2})$, which we have agreed to ignore. Locally nonrotating
[42] particles will have an angular velocity $\frac{d\phi }{dt}=-g_{t\phi
}/g_{\phi \phi }=\frac{\omega (r)}{\sin ^{2}\theta }$ but again the order of
the effect would go like $\omega \times C\sim O(\epsilon ^{2})$. Thus the CS
effect can appear only in $O(\epsilon ^{2})$. This is a bit unfortunate. The
spin precession effect, as discussed by KMAT, seems better suited to study
the CS effect.

\begin{center}
\textbf{VII. Summary}
\end{center}

As shown by KMAT [33], the constant parameter $C_{2}$ in their solution is
the observed (roughly constant) tangential velocity of particles moving in
equatorial circular orbits in the galactic halo. Our aim here was to
theoretically investigate the influence of such CS parameters on the quantum
phase shift in thermal neutron interferometry. \ To achieve this goal,
several steps were needed. We first established the stability of circular
orbits in the CS theory, which is essential for the physical viability of
the model itself. The second step was to preview Anandan's arguments. In the
third step, we familiarized the readers with the notion of curved spacetime
optical-mechanical analogy, especially the reformulation of geodesic
equations developed by Alsing [36]. (We relax the assumption that KMT\
solution (6) replaces that of Kerr but there are indications that the CS
modification is a non-trivial perturbation of Kerr geometry [35]). In the
final step, we applied these equations in the linear aproximation to obtain
the effects on the phase shift contributed by inertial forces, special
relativity, general relativity and the CS modified theory.

Eq.(69) shows that, to leading order, the CS constants influence only the
pure gravity term $\Delta \phi _{g}$. However, the constant $C_{1}$ is
observationally clueless yet, but $C_{2}$ already has a solid observable
meaning. We can set $C_{1}=0$ without any loss of rigor, since $\omega (r)$
of Eq.(7) would still satisfy the core CS equation [$r(r-2M)\omega ^{\prime
\prime }+4(r-2M)\omega ^{\prime }+2\omega =0$]. Stability of circular orbits
is not hampered when $C_{1}=0$. Hence, leaving aside relativistic terms $%
O(c_{0}^{-2})$, we have $\Delta \phi _{CSG}\simeq \left[ 1+(\alpha +\beta
)+\alpha \beta \right] \left( \Delta \widetilde{\phi }_{g}\right) +(1+\gamma
)(\Delta \widetilde{\phi }_{cor.})$, where $\alpha =\frac{4\ell C_{2}}{\rho }
$, $\beta =\frac{2M}{\rho }$, $\gamma =\frac{5M}{\rho }$, representing
respectively the CS and Schwarzschild influences on the observed flat space
effects of Newtonian gravity and inertial forces.

One might argue that the mass $M$ in question is actually that of the
central galactic bulge and any possibility of interferometry is out of
question. We wish to argue differently. Observationally, the frequency
shifts in the HI radiation show that, in the halo region, $C_{2}=v^{\varphi
}/c_{0}$ \ is nearly constant at a value $7\times 10^{-4}$ [43]. We may take
this value of $C_{2}$ as permanently fixed (up to observational
uncertainty), like cosmological constant, and treat the KMT solution as
completely determined like Schwarzschild-de Sitter solution. Then we can
apply the solution to study the nonlocal effect of this constant on the
measurement of phase shifts for \textit{any} rotating source, including
Earth. To lowest order, and assuming $\omega \simeq \frac{C_{2}}{\rho +M}$
in Eq.(42), we see that $\alpha =\frac{4g_{\phi }C_{2}}{\rho }=4C_{2}^{2}(1+%
\frac{3M}{\rho })\sim 2\times 10^{-6}$. The Schwarzschild ($\beta $, $\gamma 
$) corrections are of the order of $\frac{M}{\rho }\sim 10^{-9}$ on Earth.
Thus the current accuracy level measuring $\Delta \phi _{g}$ must be
augmented by $10^{6}$ times for CS effect and $10^{9}$ times for the
Schwarzschild effect. Whatever be the experimental scenario, our final
result, Eq.(69), provides us with a clear view of the various effects within
a single theoretical framework. It would be interesting to do analogous
calculations for Dirac particles. But it would be highly nontrivial since
the Dirac Lagrangian density would have to be coupled to the CS field,
through the covariant derivative. Perhaps a starting point could be the
analysis of Alexander and Yunes [31] but extending this to spinning
particles would be very hard.

Finally we would like to summarize the problematic characteristics of the
KMT solution it has inherited from the non-dynamical framework. First, the
solution is not a well accepted replacement of slow rotation limit of the
Kerr solution. The KMT\ solution is a very special line element that holds
only for a particular choice of the CS coupling field. Unlike in the Kerr
case, there is no Robinson's theorem [44] for rotating sources in CS gravity
that could suggest that this solution is unique. Indeed, Yunes and Pretorius
[35] have found a family of new stationary axisymmetric solutions in the
non-dynamical theory, which shows that solutions in this theory can not be
unique. From this standpoint, the far-field solutions of Alexander and
Yunes, and Smith \textit{et al} are equally valid. Second, the KMT solution
is incompatible with the dynamical formulation of CS modified gravity. This
is because the energy contained in the KMT CS scalar is infinite. The
Yunes-Pretorius solution bypasses the problem with the stress tensor and is
compatible with dynamical theory. Third, the non-dynamical framework (upon
which both the KMT solution [34] and the far-field solutions of Alexander
and Yunes [27, 28] are based) is problematic for Schwarzschild black hole
perturbation theory, as shown by Yunes and Sopuerta [32]. In this work, the
authors showed that there cannot be single-parity perturbations in
non-dynamical CS modified gravity, which suggests that when a small object
falls head-on into a supermassive black hole, the latter cannot be perturbed
to linear order. This obstacle is partially resolved in the dynamical
theory. Fourth, the non-dynamical framework is also problematic from the
theoretical standpoint. The main objection comes from the external, purely 
\textit{ad hoc} prescription of the CS coupling field, which is not
motivated by considerations from within the theory. Moreover, the
non-dynamical framework also contains an additional constraint (the
Pontryagin constraint), which suggests that the modified field equations are
over-constrained. Indeed, this constraint affects the black hole
perturbation theory (Yunes and Sopuerta [32]), and the search for exact
solutions that could represent spinning black holes (Grumiller and Yunes
[30]). If this is the case, then one might conclude that the non-dynamical
framework can only have an ill-posed initial value problem [32].

\bigskip

\textbf{Figure captions:}

Fig,1 The height represents $V^{\prime \prime }$ for small values of $%
C_{1}\neq 0$, $C_{2}\neq 0$ along x and y axes respectively. The radius is
arbitrary but fixed at any $r=R$. Negative values of $V^{\prime \prime }$
indicate stability.

Fig.2. The de Broglie wavelets originating at $A$ and interfering at $B$,
one travelling along $ADB$ and the other along $ACB$, the plane $ADBC$ being
rotated about the horizontal axis $AO$ (Ref.[14]).

\bigskip

\textbf{Acknowledgments:}

This paper is dedicated to Professor Swapan K. Ghosal on his 62$^{\text{nd}}$
birthday. We are deeply indebted to Professor Alexander I. Filippov for many
stimulating discussions and to Guzel N. Kutdusova, Denis V. Kondratiev,
Aydar R. Bikmetov and Rustam R. Fazlaev for their interest and useful
assistance. We sincerely thank the anonymous referee for many incisive and
informative comments. KKN acknowledges warm hospitality at BSPU and SSPA.

\bigskip

\textbf{References}

[1] The literature is too vast. However, for a review on the state of
research in quantum gravity, see: C. Rovelli, \textit{Living Reviews in
Relativity}, vol.1 (http://www.livingreviews.org/Article).

[2] R. Colella, A.W. Overhauser, and S.A. Werner, Phys. Rev. Lett. \textbf{34%
}, 1472 (1975); A.W. Overhauser and R. Colella, Phys. Rev. Lett. \textbf{33}%
, 1237 (1974); D. Greenberger and A.W. Overhauser, Rev. Mod. Phys. \textbf{51%
}, 43 (1979).

[3] S.A. Werner, J.-L. Staudenmann, and R. Colella, Phys. Rev. Lett. \textbf{%
42}, 1103 (1979); T.L. Gustavson, P. Bouyer, and M.A. Kasevich, Phys. Rev.
Lett. \textbf{78}, 2046 (1997).

[4] Lorne A. Page, Phys. Rev. Lett. \textbf{35}, 543 (1975).

[5] T. Padmanabhan, T.R. Seshadri, and T.P. Singh, Int. J. Mod. Phys. A 
\textbf{1}, 491 (1986); V. Daftardar and N. Dadhich, Int. J. Mod. Phys. A 
\textbf{2}, 731 (1987); K.K. Nandi, A. Islam, and J. Evans, Int. J. Mod.
Phys. A \textbf{12}, 3171 (1997).

[6] Abhay Ashtekar and Anne Magnon, J. Math. Phys. \textbf{16}, 342 (1975);
N. Fornengo, C. Giunti, C.W. Kim, and J. Song, Phys. Rev. D \textbf{56},
1895 (1997); C. Giunti, C.W. Kim, and U.W. Lee, Phys. Rev. D \textbf{44},
3635 (1991); S. Capozziello and G. Lambiase, Mod. Phys. Lett. A \textbf{14},
2193 (1999).

[7] D.V. Ahluwalia and C. Burgard, Phys. Rev. D \textbf{57}, 4724 (1998);
D.V. Ahluwalia and C. Burgard, Gen. Rel. Grav. \textbf{28}, 1161 (1996); J.
Wudka, Mod. Phys. Lett. A \textbf{6}, 3291 (1991).

[8] T. Bhattacharya, S. Habib, and E. Mottola, Phys. Rev. D \textbf{59},
067301 (1999).

[9] M. Gasperini, Phys. Rev. D \textbf{39}, 3606 (1989); A. Halprin and C.N.
Leung, Phys. Rev. Lett. \textbf{67}, 1833 (1991).

[10] L. Stodolsky, Gen. Rel. Grav. \textbf{11}, 391 (1979).

[11] P.M. Alsing, J.C. Evans, and K.K. Nandi, Gen. Rel. Grav. \textbf{33},
1459 (2001).

[12] Y. Fukuda \textit{et al}., Phys. Rev. Lett. \textbf{82}, 1810 (1998).

[13] G. Amelino-Camelia, J. Ellis, N.E. Mavromatos, D.V. Nanopoulos, and S.
Sarkar, Nature (London), \textbf{393}, 763 (1998).

[14] J. Anandan, Phys. Rev. D \textbf{15}, 1448 (1977); J. Anandan, Nuovo
Cim. A \textbf{53}, 221 (1979).

[15] K.K. Nandi and A. Islam, Am. J. Phys. \textbf{63}, 251 (1995)

[16] J. Evans, A. Islam, and K.K. Nandi, Am. J. Phys. \textbf{64}, 1404
(1996)

[17] J. Evans, A. Islam, and K.K. Nandi, Gen. Rel. Grav. \textbf{28}, 413
(1996)

[18] J. Evans, P.M. Alsing, S. Giorgetti, and K.K. Nandi, Am. J. Phys. 
\textbf{69}, 1103 (2001).

[19] J. Evans and M. Rosenquist, Am. J. Phys. \textbf{54}, 876 (1986).

[20] K.K. Nandi and Y.Z. Zhang, Phys. Rev. D \textbf{66}, 063005 (2002).

[21] M.B. Green and J.H. Schwarz, Phys. Lett. \textbf{149 B}, 117 (1984); A.
Sen, Phys. Rev. Lett. \textbf{69}, 1006 (1992).

[22] J. Polchinski, \textit{String theory}, Cambridge University Press,
Cambridge (1998).

[23] S.H.S. Alexander, M.E. Peshkin, and M.M. Sheikh-Jabbari, Phys. Rev.
Lett. \textbf{96}, 081301 (2006); \textit{eprint}:hep-th/0701139.

[24] S.H.S. Alexander and S.J. Gates Jr., J. Cosmol. Astropart. Phys. 06
(2006) 018.

[25] R. Jackiw and S.-Y. Pi, Phys. Rev. D \textbf{68}, 104012 (2003).

[26] D. Guarrera and A.J. Hariton, Phys. Rev. D \textbf{76}, 044011 (2007).

[27] S.H.S. Alexander and N. Yunes, Phys. Rev. Lett. \textbf{99}, 241101
(2007).

[28] S.H.S. Alexander and N. Yunes, Phys. Rev. D \textbf{75}, 124022 (2007).

[29] T. L. Smith, A. L. Erickcek, R. R. Caldwell, and M. Kamionkowski, Phys.
Rev. D \textbf{77}, 024015 (2008).

[30] D. Grumiller and N. Yunes, Phys. Rev. D \textbf{77}, 044015 (2008).

[31] S.H.S. Alexander and N. Yunes, Phys. Rev. D \textbf{77}, 124040 (2008).

[32] N.Yunes and C.F. Sopuerta, Phys. Rev. D \textbf{77}, 064007 (2008).

[33] K. Konno, T. Matsuyama, Y. Asano, and S. Tanda, Phys. Rev. D \textbf{78}%
, 024037 (2008).

[34] K. Konno, T. Matsuyama, and S. Tanda, Phys. Rev. D \textbf{76}, 024009
(2007).

[35] N. Yunes and F. Pretorius, \textit{eprint}: gr-qc/0902.4669.

[36] P.M. Alsing, Am. J. Phys. \textbf{66}, 779 (1998).

[37] M. Mathisson, Acta Phys. Polon. \textbf{6}, 163 (1937); A. Papapetrou,
Proc. Roy. Soc. London, A \textbf{209}, 258 (1951).

[38] L.D. Landau and E.M. Lifschitz, \textit{The Classical Theory of Fields}%
, Pergamon, New York (1975), 4th Ed.

[39] K.S. Thorne, R.H. Price, and D. MacDonald, \textit{Black Holes:\ The
Membrane Paradigm}, Yale University Press, New Haven (1986).

[40] E. Harris, Am. J. Phys. \textbf{59}, 421 (1991).

[41] In the Kerr case, $g_{\phi }=-\frac{2Ma}{\rho }$. Solving for $\frac{%
d\phi }{dt}$, using the weak field approximation, $\rho =r-M$, it follows
that $\ell =0$ leads to $\rho ^{2}\frac{d\phi }{dt}=\frac{2Ma}{\rho }$. It
means that the test particle coming straight in from infinity acquires an
angular velocity $\frac{2Ma}{\rho ^{3}}$ in the same sense as the rotation
of $M$. This is essentially the Lense-Thirring frame dragging effect.

[42] C.W. Misner, K.S. Thorne and J.A. Wheeler, \textit{Gravitation},
Freeman, San Francisco (1973), p. 895.

[43] J.J. Binney and S. Tremaine, \textit{Galactic Dynamics}, Princeton
University Press, New Jersey (1987); A. Borriello and P. Salucci, Mon. Not.
R. Astron. Soc. \textbf{323}, 285 (2001); M. Persic, P. Salucci, and F.
Stel, Mon. Not. R. Astron. Soc.\textit{\ }\textbf{281}, 27 (1996).

[44] D.C. Robinson, Phys. Rev. Lett. \textbf{34}, 905 (1975).

\bigskip

\end{document}